\documentclass[english,reprint,prd,aps,showpacs]{revtex4-1}

\usepackage[T1]{fontenc}
\usepackage[latin1]{inputenc}
\usepackage{amsmath}
\usepackage{graphicx}
\usepackage{amssymb}
\usepackage{esint}

\usepackage{longtable}
\usepackage{dcolumn}
\usepackage{babel}

\makeatother

\usepackage{babel}

\makeatletter

\begin{document}

\title{Wormholes in de Sitter branes}

\author{C. Molina}
\email{cmolina@usp.br}
\affiliation{Escola de Artes, Ci\^{e}ncias e Humanidades, Universidade de
  S\~{a}o Paulo\\ Av. Arlindo Bettio 1000, CEP 03828-000, S\~{a}o
  Paulo-SP, Brazil}

\author{J. C. S. Neves}
\email{juliano@fma.if.usp.br}
\affiliation{Instituto de F\'{\i}sica, Universidade de S\~{a}o Paulo \\
C.P. 66318, 05315-970, S\~{a}o Paulo-SP, Brazil}

\begin{abstract}
In this work, we present a class of geometries which describes wormholes in a Randall-Sundrum brane model, focusing on de Sitter backgrounds.  Maximal extensions of the solutions are constructed and their causal structures are discussed. A perturbative analysis is developed, where matter and gravitational perturbations are studied. Analytical results for the quasinormal spectra are obtained and an extensive numerical survey is conducted. Our results indicate that the wormhole geometries presented are stable.
\end{abstract}

\pacs{04.20.Jb,04.50.Gh,04.50.Kd}



\maketitle

\section{Introduction}

Wormholes are compact space-times with nontrivially topological interiors and topologically simple boundaries. They can be seen as connections between different universes or topological handles between distant parts of the same universe. Although they are certainly exotic structures, they appear as exact solutions of Einstein equations with physically relevant scenarios and are compatible with the usual local physics \cite{Visser-livro}. Samples of the work developed include solutions in usual general relativity \cite{Visser-livro,Morris_Thorne,Morris_Thorne_Yurtsever,Lemos-2003,Molina-2011}, in Gauss-Bonnet theory \cite{Bhawal,Dotti_Trancoso}, in Brans-Dicke theory \cite{Agnese,Anchordoqui,Nandi1,Nandi2,Lobo}, and brane world context \cite{Casadio,Bronnikov2,Bronnikov,Lobo2}.

One motivation in the treatment of wormhole physics was due to the results of Morris, Thorne, and Yurtsever \cite{Morris_Thorne,Morris_Thorne_Yurtsever}, which connected time machines and traversable wormholes. More recently, new cosmological observations and theoretical proposals have motivated a renewed interest in geometries which describe Lorentzian wormholes. One of their general characteristics is that wormholes must be supported by ``exotic matter,'' which violates usual energy conditions. Nevertheless, recent observations suggest that the Universe may be dominated by some form of exotic matter \cite{Riess-Perlmutter,Caldwell}, which make wormhole scenarios more plausible. Other source of geometries with nontrivial topology are brane worlds. In this context, the wormhole is supported by the influence of a bulk in the brane which describes our Universe. It is in this framework that the present paper is inserted.

In this work we derive a family of asymptotically de Sitter wormhole solutions in a brane world context, more specifically in a Randall-Sundrum-type model \cite{Randall-Sundrum}. We used the effective gravitational field equations derived by Shiromizu, Maeda, and Sasaki \cite{Shiromizu}. As there are few satisfactory bulk solutions for compact objects in Randall-Sundrum scenarios, one alternative is to build geometries in the brane and invoke Campbell-Magaard theorems \cite{Seahra}, which guarantees their extensions through the bulk (locally at least). This approach has been used by several authors \cite{Dadhich_Maartens,Casadio,Bronnikov,Bronnikov2,Lobo2}, and we will be following it in the present work. Moreover, global regularity of the brane is expected to facilitate the construction of a regular bulk solution \cite{Bronnikov2}. This issue will be explored in the space-times constructed here.

Contrary to what has been suggested in the literature \cite{Bronnikov,Molina_Neves}, we obtain de Sitter solutions which are regular everywhere. The class of geometries studied here complement the asymptotically flat space-times treated in  \cite{Dadhich_Maartens,Casadio,Bronnikov,Bronnikov2,Lobo2}; and the asymptotically anti-de Sitter metrics in \cite{Lemos-2004,Barcelo,Molina_Neves}. We should mention that solutions of the effective Einstein equations with positive cosmological constant in a brane setting were previously considered in \cite{Lobo2}. While there is some overlap between the present work and \cite{Lobo2}, we have explored some global issues not considered in the mentioned paper, such as regularity and the existence of cosmological horizons. As will be discussed, these points are particularly important for de Sitter geometries.

If one considers the possibility of the existence of wormholes seriously, characteristics such as stability and response of wormholes to external perturbations should be investigated. Perturbative dynamics around wormholes \cite{wh-perturbations} have not been as thoroughly explored as the black hole problem. We further advance the perturbative treatment of wormhole geometries in the present work. Matter and gravitational perturbations are considered in the background of the de Sitter wormhole geometries derived here. 

The structure of this paper is presented in the following. In Sec. II we have derived a family of analytic asymptotically de Sitter solutions in an Randall-Sundrum-type brane. In Sec. III the maximal extensions of the solutions are considered and the wormhole geometries discussed. The near extreme limit of the wormhole solutions are considered in Sec. IV. Section V deals with the perturbative analysis of the backgrounds derived. And finally in Sec. VI some final remarks are made. In this work we have used the metric signature $diag(-+++)$ and the geometric units $G_{4D}=c=1$, where $G_{4D}$ is the effective four-dimensional gravitational constant.

\section{de Sitter brane solutions}

The basic brane world set up is a four-dimensional brane, our universe, immersed in a larger manifold, the bulk. It is generally postulated that the usual matter fields are confined in the brane \cite{Maartens}. Following the approach suggested by Shiromizu, Maeda and Sasaki \cite{Shiromizu}, the effective four-dimensional gravitational field equations in the a vacuum Randall-Sundrum brane is 
\begin{equation}
R_{\mu\nu} - \frac{1}{2} R g_{\mu\nu} = -\Lambda_{4D} g_{\mu\nu} - E_{\mu\nu} \,\, .
\label{eq_projetada}
\end{equation}
In this effective Einstein equation,  $\Lambda_{4D}$ is the four-dimensional
brane cosmological constant and $E_{\mu\nu}$ is proportional to
the (traceless) projection on the brane of the five-dimensional Weyl
tensor. Eqs. (\ref{eq_projetada}) reduce to usual four-dimensional vacuum Einstein equations in the low-energy limit.

If we impose staticity and spherical symmetry in the brane, that is,
\begin{equation}
ds^{2} = -A(r)dt^{2} + \frac{dr^{2}}{B(r)} + r^{2} (d\theta^{2} + \sin^{2} \theta d\phi^{2}) \,\, ,
\label{metrica_4d}
\end{equation}
the trace of Eq. (\ref{eq_projetada}) will be 
\begin{equation}
R = 4\Lambda_{4D} \,\, ,
\label{Ricci_scalar}
\end{equation}
where $R$ denotes the four-dimensional Ricci scalar. The Eq.(\ref{Ricci_scalar}) may be written as a constraint between the functions $A$ and $B$ 
\begin{gather}
2(1-B) - r^{2}B \left\{ \frac{A''}{A} - \frac{(A')^{2}}{2A^{2}} + \frac{A'B'}{2AB} + \frac{2}{r} \left[\frac{A'}{A} + \frac{B'}{B}\right]\right\} \nonumber \\
= 4\Lambda_{4D} \,\, ,
\label{constraint}
\end{gather}
with prime ($'$) denoting differentiation with respect to $r$.

We propose to construct asymptotically de Sitter space-times, such
that $\Lambda_{4D} > 0$. In addition, we assume that they are ``close''
to the usual spherically symmetric (electro)vacuum solution given by $A=A_{0}$ and $B=B_{0}$, with
\begin{equation}
A_{0}(r) = B_{0}(r) = 1 - \frac{2M}{r} + \frac{Q^{2}}{r^{2}} - \frac{\Lambda_{4D}}{3}r^{2} \,\, ,
\label{vac_solution}
\end{equation}
where $M$ and $Q^{2}$ are positive constants. Denoting a particular
solution by the pair $(A,B)$ of functions which satisfy the constraint
(\ref{constraint}), we are searching for a family of solutions $\mathcal{S}$
such that: 

(i) the vacuum solution $(A_{0},B_{0})$ is an element of $\mathcal{S}$;

(ii) a generic solution $(A_{C_{1}},B_{C_{1}})\in\mathcal{S}$ is
a continuous deformation of $(A_{0},B_{0})$, that is, there is (at
least) one set of solutions $\mathcal{D}_{C_{1}} = \left\{ \left(A_{C},B_{C}\right),0\leq C \leq C_{1}\right\} $, labeled by a real parameter $C$, such that $\mathcal{D}_{C_{1}} \subset \mathcal{S}$.

Since Eq.(\ref{constraint}) is linear in terms of $B$, a linear
combination of solutions with $A$ fixed is still a solution. Moreover,
since we are interested in deformations of the usual vacuum solutions,
we assume the \emph{Ansatz}
\begin{equation}
A(r) = A_{0}(r) \,\, ,
\label{ansatz_A}
\end{equation}
\begin{equation}
B(r) = B_{0}(r) - C \, B_{lin}(r) \,\, ,
\label{ansatz_B}
\end{equation}
with $\partial B_{lin}/\partial C=0$. Using Eqs.(\ref{ansatz_A})
and (\ref{ansatz_B}), the constraint (\ref{constraint}) can be rewritten
as a linear first order ordinary differential equation on the $B_{lin}$
\begin{equation}
h(r) \, \frac{dB_{lin}(r)}{dr} + f(r) \, B_{lin}(r) = 0,
\label{eq_Blin}
\end{equation}
where the functions $h$ and $f$ are given by
\begin{equation}
h(r) = 4A_{0} + r A_{0}' \,\, ,
\label{function_h}
\end{equation}
\begin{equation}
f(r) = \frac{4A_{0}}{r} + 4A_{0}' + 2rA_{0}'' - \frac{r\,\left(A_{0}'\right)^{2}}{A_{0}} \,\, .
\label{function_f}
\end{equation}
 
The zero structure of $h$ and $A_{0}$ will be of great importance.
If $M>0$, $Q \ne 0$ and $0 < \Lambda_{4D} < \Lambda_{ext}$, where 
\begin{gather}
\Lambda_{ext} =  \nonumber \\
\frac{3}{8Q^{2}} - \frac{1}{32} \left[ \left(\frac{9M^{2}}{Q^{3}} - \frac{6}{Q} \right)^{2} -3M\left( \frac{9M^{2}}{Q^{4}} - \frac{8}{Q^{2}}\right)^{3/2} \right]
\label{Lambda_critico}
\end{gather}
is the critical value of $\Lambda_{4D}$, the function $A_{0}$
has four real zeros $r_{c}$, $r_{+}$, $r_{-}$ and $r_{n}$ such
that $r_{n}<0<r_{-}<r_{+}<r_{c}$. Also in this region of the parameter
space, the function $h$ has four real zeros $r_{0}$, $r_{0-}$,
$r_{0--}$ and $r_{0n}$ with $r_{0n}<0<r_{0--}<r_{0-}<r_{0}$. Explicit
expressions for the several roots introduced are straightforward but
cumbersome. Of fundamental importance in this work is the relation $r_{+}<r_{0}<r_{c}$, which is always satisfied for $0 < \Lambda_{4D} < \Lambda_{ext}$.

The solution of (\ref{eq_Blin}) for the correction $B_{lin}$,
general up to a multiplicative integration constant, is given by
\begin{equation}
B_{lin}(r) = A_{0}(r) \, \frac{\left( r - r_{0--} \right)^{c_{0--}}}{\left(r - r_{0}\right)^{c_{0}}\left(r - r_{0-}\right)^{c_{0-}}\left(r - r_{0n}\right)^{c_{0n}}} \,\, ,
\label{Blin}
\end{equation}
where the positive constants $c_{0}$, $c_{0-}$, $c_{0--}$ and $c_{0n}$ are
\begin{equation}
c_{0} = \frac{2}{\Lambda_{4D}} \, \frac{r_{0}\left(2\Lambda_{4D}r_{0}^{2} - 1\right)}{\left(r_{0} - r_{0-}\right)\left(r_{0} - r_{0--}\right)\left(r_{0} - r_{0n}\right)} \,\, ,
\end{equation}
\begin{equation}
c_{0-} = - \frac{2}{\Lambda_{4D}} \, \frac{r_{0-}\left(2\Lambda_{4D}r_{0-}^{2} - 1\right)}{\left(r_{0} - r_{0-}\right)\left(r_{0-} - r_{0--}\right)\left(r_{0-} - r_{0n}\right)} \,\, ,
\end{equation}
\begin{equation}
c_{0--} = - \frac{2}{\Lambda_{4D}} \, \frac{r_{0--}\left(2\Lambda_{4D}r_{0--}^{2} - 1\right)}{\left(r_{0} - r_{0--}\right)\left(r_{0-} - r_{0--}\right)\left(r_{0--} - r_{0n}\right)} \,\, ,
\end{equation}
\begin{equation}
c_{0n} = -\frac{2}{\Lambda_{4D}} \, \frac{r_{0n}\left(2\Lambda_{4D}r_{0n}^{2} - 1\right)}{\left(r_{0--} - r_{0n}\right)\left(r_{0} - r_{0n}\right)\left(r_{0-} - r_{0n}\right)} \,\, .
\end{equation}
Therefore the complete solutions for $A$ and $B$ can be expressed as
\begin{equation}
A(r) = A_{0}(r) = \frac{\Lambda_{4D}}{3r^{2}} \left(r_{c} - r\right) \left(r - r_{+}\right) \left(r - r_{-}\right) \left(r - r_{n}\right) \,\, ,
\label{sol_A}
\end{equation}
\begin{equation}
B(r) = A_{0}(r) \left[1 - \frac{C}{\left(r - r_{0}\right)^{c_{0}}} \, \frac{\left(r - r_{0--}\right)^{c_{0--}}}{\left(r - r_{0-}\right)^{c_{0-}} \left(r - r_{0n}\right)^{c_{0n}}}\right] \,\, .
\label{sol_B}
\end{equation}

It is apparent that the function $B$ diverges in the limit $r\rightarrow r_{0}$.
Since $r_{+}<r<r_{c}$ is a natural candidate for the space-time static
region and $r_{+}<r_{0}<r_{c}$, previous works in the literature \cite{Bronnikov,Molina_Neves} have suggested that regular de Sitter solutions of (\ref{Ricci_scalar}) might not exist. However, we will show that this is not so. 

As will be discussed in the following sections, the main characteristics
of this class of solutions are captured by the simpler case $M=Q=0$.
In this limit the coefficients $r_{c}$ and $r_{0}$ can be easily
expressed as
\begin{equation}
r_{c} = \sqrt{\frac{3}{\Lambda_{4D}}} \,\, , \,\, r_{0} = \sqrt{\frac{2}{\Lambda_{4D}}} \,\, ,
\end{equation}
and $r_{n} = -r_{0}$, $r_{0n} = -r_{0}$, $c_{0} =  c_{0n} = 3/2$, $c_{0-} = 1$. The
remaining constants $r_{-}$, $r_{--}$, $r_{0-}$, $r_{0--}$ and
$c_{0--}$ are null. The metric functions are given by
\begin{equation}
A(r) = 1 - \frac{r^{2}}{r_{c}^{2}} \,\, ,
\label{A_M_zero}
\end{equation}
\begin{equation}
B(r) = \left(1 - \frac{r^{2}}{r_{c}^{2}}\right) \left[1 - C\,\frac{1}{r\,\left(r^{2} - r_{0}^{2}\right)^{3/2}}\right] \,\, .
\label{B_M_zero}
\end{equation}

The energy density, radial and tangential pressures associated with
Eqs.(\ref{A_M_zero}) and (\ref{B_M_zero}) may be defined as
\begin{equation}
\left(-E_{\mu}^{\nu}\right) = 8\pi
\left(\begin{array}{cccc}
 -\rho\\
 & p_{r}\\
 &  & p_{t}\\
 &  &  & p_{t} \end{array}\right)
\end{equation}
and are given by 
\begin{equation}
8\pi \, \rho = \frac{C}{3r_{0}^{2}} \, \frac{2r^{2} - 5r_{0}^{2}}{r \left(r^{2} - r_{0}^{2}\right)^{5/2}} \,\, ,
\end{equation}
\begin{equation}
8\pi\, p_{r} = \frac{C}{r_{0}^{2}} \, \frac{2r^{2} - r_{0}^{2}}{r^{3} \left(r^{2} - r_{0}^{2}\right)^{3/2}} \,\, ,
\end{equation}
\begin{equation}
8 \pi \, p_{t} = -\frac{C}{6r_{0}^{2}} \, \frac{4r^{4} - 4r_{0}^{2}r^{2} + 3r_{0}^{4}}{r^{3} \left(r^{2} - r_{0}^{2}\right)^{5/2}} \,\, . 
\end{equation}

\pagebreak

\noindent
These energy density and pressures are not generally positive-definite,
and the effective stress-energy tensor $\left(-E_{\mu}^{\nu}\right)$
do not satisfy usual energy conditions. Still, in the context of this
work, they should be viewed as \emph{effective quantities}, associated
with a \emph{vacuum} brane model.

\section{Wormholes inside cosmological horizons}

Strictly speaking, the metric described by the functions $A$ and $B$ in Eqs.(\ref{sol_A}) and (\ref{sol_B}), or in Eqs.(\ref{A_M_zero}) and (\ref{B_M_zero}), describes a space-time only for the values of the radial parameter $r$ such that $A(r)>0$ and $B(r)>0$. The maximal extensions of these solutions will be presently treated. At this point, an important question to be treated is the range of the parameters for which the solution given by Eqs.(\ref{sol_A}) and (\ref{sol_B}) describes an acceptable geometry. 

If $C<0$, the functions $A$ and $B$ are positive for $r_{+}<r<r_{c}$. But $r_{+}<r_{0}<r_{c}$, so $B$ in Eq.(\ref{sol_B}) is divergent at $r_{0}$. The geometry is well-defined and static for $r>r_{0}$, but its curvature invariants are not bounded, as seen by the behavior of the Kretschmann scalar near $r_{0}$ 
\begin{equation}
\lim_{r\rightarrow r_{0}} \left| R_{\alpha\beta\gamma\delta}R^{\alpha\beta\gamma\delta} \right| \rightarrow \infty \,\, .
\end{equation}
Therefore, for this case \emph{a naked curvature singularity} is present at $r\rightarrow r_{0}$. This solution will not be further explored in the present work.

If $C=0$, we recover the usual Reissner-Nordstr\"{o}m-de Sitter vacuum solution, and the regular region is given by $r_{+}<r<r_{c}$. As is well-known, in nonextremal regimes the surfaces $r=r_{+}$ and $r=r_{c}$ describe an event and an cosmological horizons in the maximal extension, respectively. One interpretation for this result is that, although the solutions with $C\ne0$ and the Reissner-Nordstr\"{o}m-de Sitter black holes have very different global characteristics, they nevertheless are locally arbitrarily close.

If $C>0$, the function $B$ is not positive-definite between $r_{+}$ and $r_{c}$. It has a third zero at $r=r_{thr}$. The relevant point is 
\begin{equation}
r_{+} < r_{0} < r_{thr} < r_{c} \,\, ,
\end{equation}
with the
functions $A$ and $B$ positive-definite and analytic for $r_{thr}<r<r_{c}$.
Therefore, the chart $(t,r,\theta,\phi)$ is valid in the region $r_{thr}<r<r_{c}$.
The analytic extension beyond $r=r_{thr}$ is suggested with the use
of the proper length $\ell$ as radial function, where 
\begin{equation}
\frac{d\ell(r)}{dr} = \frac{1}{\sqrt{B(r)}} \,\, . 
\label{proper_length}
\end{equation}
Choosing an appropriate integration constant in Eq.(\ref{proper_length}),
the region $r_{thr}<r<r_{c}$ is mapped into $0<\ell<\ell_{max}$,
with a finite $\ell_{max}$. The extension is made analytic continuing
the metric with $-\ell_{max}<\ell<\ell_{max}$. The resulting geometry
has a wormhole structure, with a throat at $r=r_{thr}$. 

The extension beyond $r=r_{c}$ can be made, for example, with
the ingoing and outgoing Eddington charts $(u,t,\theta,\phi)$ and
$(v,t,\theta,\phi)$, where $u$, $v$ are the light-cone variables
\begin{equation}
u = t - r_{\star} \,\,\, \textrm{and} \,\,\, v = t + r_{\star} \,\, .
\label{u_v_definitions}
\end{equation}
The radial variable $r_{\star}$ is the tortoise coordinate, defined as
\begin{equation}
\frac{dr_{\star}(r)}{dr} = \frac{1}{\sqrt{A(r)\: B(r)}} \,\, .
\label{tortoise_coordinate}
\end{equation}
In the maximal extension, the surface $\ell = \ell_{max}$ ($r=r_{c}$)
is a Killing horizon, interpreted as a cosmological horizon. A more
physical interpretation of the geometry is a spherically symmetric
wormhole inside an exponentially expanding Universe.

Charts based on the tortoise coordinate $r_{\star}$ or the $u-v$
coordinates can be used to cover all the static region. In this case,
$\left\{ \left(t,\ell,\theta,\phi\right),\ell \in (-\ell_{max},+\ell_{max})\right\} $
is mapped into $\left\{ \left(t,r_{\star},\theta,\phi\right),r_{\star} \in (-\infty,+\infty)\right\} $
or $\left\{ \left(u,v,\theta,\phi\right),u \in (-\infty,+\infty),v\in(-\infty,+\infty)\right\} $. These coordinate systems will be used in the perturbative
analysis of the wormhole.

\begin{center}
\begin{figure}[tp]
\includegraphics[clip,width=0.5\columnwidth]{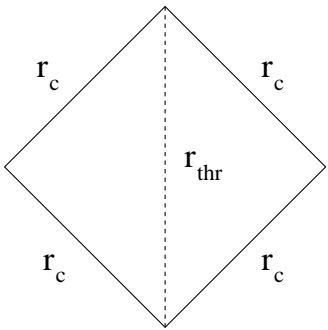} 
\caption{Conformal diagram of wormhole the solution inside a cosmological horizon. Dashed line denotes the wormhole throat.}
\label{de_sitter_diagram} 
\end{figure}
\end{center}

Applying standard procedures (see, for example, \cite{Walker}), the
Penrose diagram of wormhole geometry can be obtained. This diagram
is present in Fig.\ref{de_sitter_diagram}. 

\section{The near extreme limit}

One limit where the geometry is simpler and its perturbative analysis
is much easier (as we will see in the next section) is the near extreme
regime. We will treat this limit in the present section. 

In order that the background characterized by the metric functions
(\ref{sol_A}) and (\ref{sol_B}) describes a wormhole, the real parameter
$C$ cannot be arbitrarily large. As $C$ grows, $r_{thr}$ approaches
$r_{c}$. The extreme value for $C$ ($C_{ext}$), such that $\lim_{C\rightarrow C_{ext}}r_{thr}\rightarrow r_{c}$ with $0<C<C_{ext}$, is given by
\begin{equation}
C_{ext} = \frac{\left(r_{c} - r_{0}\right)^{c_{0}} \left(r_{c} - r_{0-}\right)^{c_{0-}} \left(r_{c} - r_{0n}\right)^{c_{0n}}}{\left(r_{c}-r_{0--}\right)^{c_{0--}}} \,\, .
\label{Cext}
\end{equation}
We will consider in this section the \emph{near extreme limit case},
where $C \lesssim C_{ext}$, that is, $C$ is very close (but still
smaller) to the maximum value $C_{ext}$. So, it is natural to define
the dimensionless parameter
\begin{equation}
\delta = \frac{r_{c}-r_{thr}}{r_{c}-r_{0}} \,\, .
\label{delta_definition}
\end{equation}
With this definition, $0<\delta<1$, since $r_{0}<r_{thr}<r_{c}$.
The near extreme regime can be characterized in terms of $\delta$
as the limit $0<\delta\ll1$. In fact, it can be shown that $C/C_{ext}=1-\mathcal{O}(\delta)$.

In the near extreme limit, the metric functions $A$ and $B$ can
approximated by the linear and quadratic functions $A^{n-ext}$ and
$B^{n-ext}$ respectively,
\begin{equation}
A(r) \approx A^{n-ext}(r) = A_{0}^{n-ext} \, \left(r_{c} - r\right) \,\, ,
\label{A_next}
\end{equation}
\begin{equation}
B(r) \approx B^{n-ext}(r) = B_{0}^{n-ext} \, \left(r - r_{thr}\right) \left(r_{c} - r\right) \,\, ,
\label{B_next}
\end{equation}
with the positive constants $A_{0}^{n-ext}$ and $B_{0}^{n-ext}$ given by
\begin{equation}
A_{0}^{n-ext} = \frac{\Lambda_{4D}}{3r_{c}^{2}} \left(r_{c} - r_{+}\right) \left(r_{c} - r_{-}\right) \left(r_{c} - r_{n}\right) \,\, ,
\end{equation}
\begin{gather}
B_{0}^{n-ext} = \nonumber \\
A_{0}^{n-ext} \left(\frac{c_{0}}{r_{c} - r_{0}} + \frac{c_{0-}}{r_{c} - r_{0-}} - \frac{c_{0--}}{r_{c}-r_{0--}} + \frac{c_{0n}}{r_{c} - r_{0n}}\right) \,\, .
\end{gather}

It is important to stress that the causal structure of the space-time
is not modified in the near extreme limit. The geometry still describes
a wormhole inside cosmological horizons, with its Penrose diagram
shown in Fig.\ref{de_sitter_diagram}. One important geometrical quantity
is the surface gravity at the cosmological horizon. In the near extreme
limit it can be explicitly calculated in terms of the roots of $A$
and $h$
\begin{eqnarray}
\kappa_{c} & = & \frac{1}{2} \left| \left. \frac{d\sqrt{A(r) \, B(r)}}{dr} \right|_{r=r_{c}} \right| \nonumber \\
& = & \frac{1}{2} \sqrt{A_{0}^{n-ext} \, B_{0}^{n-ext} \, \left(r_{c}-r_{0}\right)} \, \delta^{1/2} \,\, .
\label{kappac_near_extreme}
\end{eqnarray}

Although the cosmological horizons may be seen as close in
the near extreme limit, this is not necessarily so. In fact, the proper
radial distance between the horizons can be arbitrarily large even
in the near extreme regime. Taking the case $M=Q=0$ for simplicity,
the maximum value for the radial proper distance ($\ell_{max}$), half
the proper distance between the two cosmological horizons, is $\ell_{max} = 3\pi\,r_{c}/2 $, 
which can be arbitrarily large as $r_{c}\rightarrow\infty$ ($\Lambda_{4D} \rightarrow 0$). The proper distance between horizons is then unbounded.

As discussed in previous and following sections, charts based on the
tortoise coordinate introduced in Eq.(\ref{tortoise_coordinate})
are very convenient for several applications. In the near extreme
regime, the metric can be explicitly written in terms of $r_{\star}$
\begin{gather}
ds^{2} = \delta A_{0}^{n-ext} \left(r_{c} - r_{0}\right) \textrm{sech}^{2} \left(\kappa_{c}r_{\star}\right) \left(-dt^{2} + dr_{\star}^{2}\right) \nonumber \\
+ \left[r_{c} - \delta\left(r_{c} - r_{0}\right) \textrm{sech}^{2} \left(\kappa_{c} r_{\star}\right) \right]^{2} d\Omega_{2}^{2} \,\, .
\label{metric_near_extreme}
\end{gather}

\section{Perturbative dynamics and stability analysis}

\subsection{General considerations for the perturbative treatment}

Once the background geometry is established, one next step is to determine
its response under small perturbations. In the lowest order, background
reaction can be ignored, and the dynamics is restricted to the matter
and gravitational perturbations in a fixed geometry. As a prototype
of matter, we will consider massless and massive scalar field, not
necessarily minimally coupled to the background. The gravitational
perturbation analysis will be limited here to the axial mode dynamics.

\begin{center}
\begin{figure}[tp]
\includegraphics[clip,width=\columnwidth]{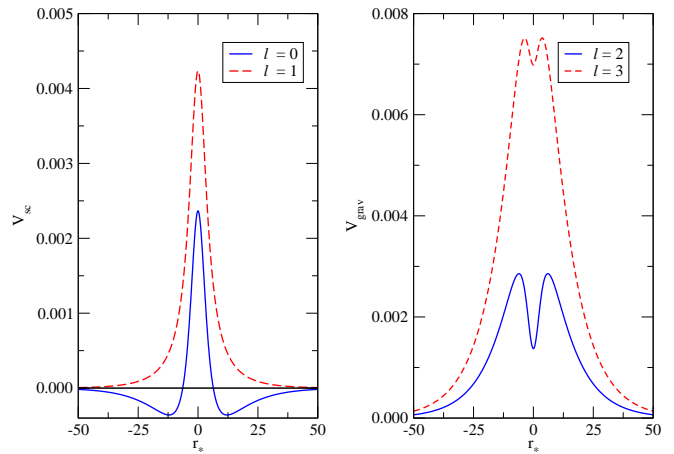} 
\caption{Scalar and gravitational effective potentials (right and left panels,
respectively) in terms of the tortoise coordinate $r_{\star}$. The
wormhole parameters used in the plots were $\Lambda_{4D}=0.01$, $M=1.0$,
$Q=0.5$ and $\delta=0.7$ ($r_{thr}=14.20$ and $r_{c}=16.23$). }
\label{effective_potentials} 
\end{figure}
\end{center}

A massless scalar perturbation field $\Phi$ is characterized by the
Klein-Gordon equation
\begin{equation}
\Box\Phi = 0 \,\, .
\end{equation}
Decomposing the scalar field $\Phi$ in terms of an expansion in spherical
harmonic components
\begin{equation}
\Phi(t,r,\theta,\phi) = \sum_{l,m} \frac{\psi_{l}(t,r)}{r} \, Y_{lm} \left(\theta,\phi\right) \,\, ,
\end{equation}
the Klein-Gordon equation give us a set of decoupled equations in
the form
\begin{equation}
-\frac{\partial^{2}\psi_{l}}{\partial t^{2}} + \frac{\partial^{2}\psi_{l}}{\partial r_{\star}^{2}} = V_{sc}(r(r_{\star})) \, \psi_{l} \,\, ,
\label{wave_equation_rstar}
\end{equation}
labeled by the multipole index $l$, with $l=0,1,2,\ldots$. The tortoise
coordinate $r_{\star}$ was introduced in Eq.(\ref{tortoise_coordinate}),
and $Y_{lm}$ denotes the spherical harmonic functions. Using results
in \cite{Abdalla-2006}, the scalar effective potential is expressed
in terms of $r$ as 
\begin{equation}
V_{sc}(r) = \frac{l(l+1)}{r^{2}}A_{0} + \frac{1}{r}A_{0}A_{0}' - \frac{C}{2r} \left(A_{0}B_{lin}\right)' \,\, .
\label{potencial_escalar}
\end{equation}
Typical profiles for the scalar effective potential are presented
in Fig.\ref{effective_potentials}.

We will consider gravitational perturbations in the brane geometry following
the treatment in \cite{Abdalla-2006}. In general, the gravitational
perturbations depend on the tidal perturbations, namely, first-order
perturbations in $E_{\mu\nu}$ ($\delta E_{\mu\nu}$). Since the complete
bulk solution is not known, we shall use the simplifying assumption
$\delta E_{\mu\nu}=0$. This assumption can be justified at least
in a regime where the energy carried in the perturbation processes
does not exceed the threshold of the Kaluza-Klein massive modes \cite{Maartens}.
Analysis of gravitational shortcuts \cite{Abdalla-2002,Abdalla-2004}
also supports this simplification, suggesting that gravitational fields
do not travel deep into the bulk. Within these premises, the gravitational
perturbation equation is
\begin{equation}
\delta R_{\mu\nu} = 0 \,\, .
\end{equation}

Following \cite{Abdalla-2006}, the gravitational axial perturbations
are given wave functions $Z_{l}$, satisfying a set of equations
of motion with the form (\ref{wave_equation_frequency}), labeled
by a multipole index $l$ ($l=2,3,\ldots$). The effective potential
in this case is given by
\begin{gather}
V_{grav}(r) = \frac{\left(l + 2\right) \left(l - 1\right)}{r^{2}}A_{0} + \frac{2}{r^{2}} A_{0}^{2} - \frac{1}{r} A_{0}A_{0}' \nonumber \\
- C\left[ \frac{2}{r^{2}} A_{0}B_{lin} - \frac{1}{2r} \left(A_{0}B_{lin}\right)^{2}\right] \,\, .
\label{potencial_gravitacional}
\end{gather}
Typical profiles are presented in Fig.\ref{effective_potentials}. 

Of particular interest in the perturbative dynamics are the so-called
quasinormal mode spectra. Consider a wave function $R$,
in the present case the scalar or gravitational perturbation ($\psi_{l}$
or $Z_{l}$), subjected to an effective potential $V$ ($V_{sc}$
or $V_{grav}$). The quasinormal modes are solutions of the ``time-independent''
version of Eq.(\ref{wave_equation_rstar}),
\begin{equation}
\frac{\partial^{2} \tilde{R}_{\omega}}{\partial r_{\star}^{2}} + \left(\omega^{2} - V\right) \tilde{R}_{\omega} = 0 \,\, ,
\label{wave_equation_frequency}
\end{equation}
satisfying both ingoing and outgoing boundary conditions asymptotically:
\begin{equation}
\lim_{r_{\star}\rightarrow\mp\infty} \tilde{R}_{\omega} \, e^{\pm i\omega r_{\star}} = 1 \,\, .
\end{equation}
The ``frequency domain'' wave function $\tilde{R}_{\omega}$ associated
with a given quasinormal mode $\omega$ is given by the Laplace transform \cite{Nollert-1992} of the function $R$ as 
\begin{equation}
\tilde{R}_{\omega}(r_{\star}) = \int_{0}^{\infty}R(t,r_{\star}) \, e^{i\omega t}\, dt \,\, .
\end{equation}
with $\omega$ extended to the complex plane. 

The relevance of the quasinormal mode calculation is twofold. They
determine the dynamical evolution of the wave function when the wave
equation is subjected to bounded initial conditions. Moreover, $\textrm{Im}\left(\omega\right)<0$
is a necessary condition for the stability
of the perturbation. The determination of quasinormal mode spectra
for the wormhole geometry will be made with analytical and numerical
techniques in the next subsections.

As will be discussed in the following, the dynamics of the perturbations
considered can be analytically treated in the near extreme limit introduced
in Sec.IV. Beyond this regime, numerical tools are necessary. In order
to analyze quasinormal mode phase and late-time behavior of the perturbations,
we apply a numerical characteristic integration scheme based in the
light-cone variables $u$ and $v$ in Eq.(\ref{u_v_definitions}),
used, for example, in \cite{Gundlach-1994,Brady-1997,Molina-2004,Konoplya-2007}.

\subsection{Spherically symmetric scalar mode ($l=0$)}

The scalar field perturbation has a spherically symmetric ($l=0$)
mode. This mode is distinct because its associated effective potential
is not positive-definite, as illustrated in Fig.\ref{effective_potentials}.
This point raises the question of whether the time evolution of the
scalar field is stable. One important result of this work is that,
in our extensive numerical investigation, the perturbation is always bounded, that is, \emph{no unstable modes were observed}. 

The presence of relevant negative peaks in the scalar potential with
$l=0$ is a potential complication for the calculation of the quasinormal frequencies. Nevertheless, the direct integration scheme used in \cite{Gundlach-1994,Brady-1997,Molina-2004,Konoplya-2007}
can be successfully employed in the present case. We will discuss
in the following some important points observed in the scalar field
evolution in the wormhole background considered.

\begin{center}
\begin{figure}[tp]
\includegraphics[clip,width=\columnwidth]{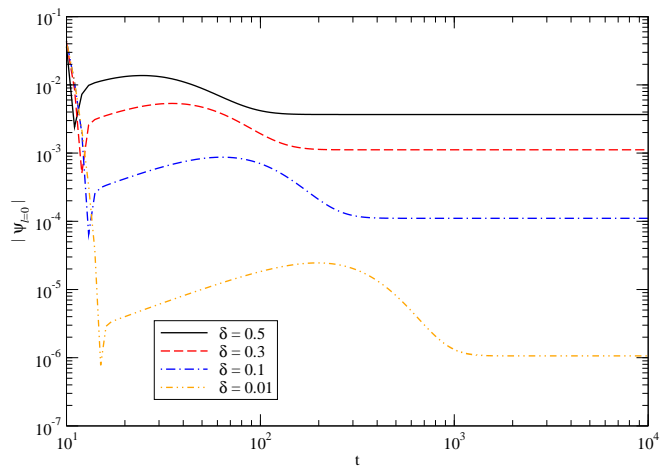} 
\caption{Scalar field evolution with $l=0$ and $r_{\star}=0$, for several
values of $\delta$. For the wormhole geometries considered, the parameters
$\Lambda_{4D}=0.01$, $M=1.0$ and $Q=0.5$ ($r_{c}=16.23$) were used. }
\label{scalar_field_evolution_l0} 
\end{figure}
\end{center}

A nonusual feature observed in the scalar dynamics is that the field
$\psi$, for a fixed value of $r_{\star}$, tends to a non-null constant
$\psi_{0}^{(0)}$ for large $t$: 
\begin{equation}
\lim_{t\rightarrow\infty} \psi_{l=0} \rightarrow \psi_{0}^{(0)} \,\, .
\end{equation}
This point is illustrated in Fig.\ref{scalar_field_evolution_l0}. A similar qualitative behavior was observed in other de Sitter geometries \cite{Molina-2004,Konoplya-2007}. 

\begin{center}
\begin{figure}[tp]
\includegraphics[clip,width=\columnwidth]{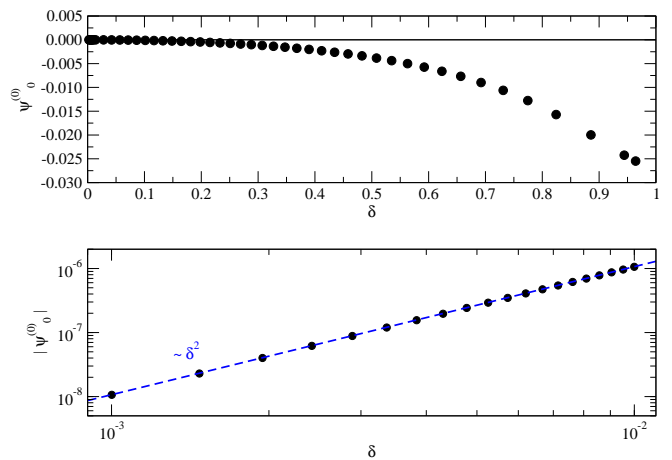}
\caption{Dependence of the asymptotic value for the $l=0$ scalar mode $\left( \psi_{0}^{(0)} \right)$ with the parameter $\delta$. The bullets indicate numerical results, and the dashed line denotes a $\delta^{2}$ power-law. For the wormhole
geometries considered, the parameters $\Lambda_{4D}=0.01$, $M=1.0$ and $Q=0.5$ ($r_{c}=16.23$) were used. }
\label{phi0_grafico} 
\end{figure}
\end{center}

In the near extreme regime, the late-time field evolution can be better
explored. The intermediate- and late-time field evolution has the form
\begin{equation}
\psi_{l=0} \simeq \psi_{0}^{(0)} + \psi_{0}^{(1)} \, e^{-\kappa_{c}t} \,\, ,
\end{equation}
with
\begin{equation}
\psi_{0}^{(0)} \propto \delta^{2} \,\, .
\end{equation}
and $\kappa_{c}$ denoting the surface gravity at the cosmological
horizon, calculated at Eq.(\ref{kappac_near_extreme}) in the near
extreme limit. The dependence of $\psi_{0}^{(0)}$ with the parameter
$\delta$ is illustrated in Fig.\ref{phi0_grafico}.

\subsection{Higher multipole modes ($l>0$) }

A general feature of the effective potentials considered when $l>0$
is that they are positive-definite. This point implies that \emph{the
dynamics is always stable for non-null} $l$. Other relevant characteristics
of both potentials are the typically complicated profiles near $r_{\star}\approx0$,
as illustrated in Fig.\ref{effective_potentials}. This latter
point makes the WKB-based methods in \cite{Schutz-1985,Iyer-1987,Konoplya-2003}
not effective in the present case, as explicitly checked by
us. The direct integration schemes used in \cite{Gundlach-1994,Brady-1997,Molina-2004,Konoplya-2007}
can still be successfully employed. Analytic results will be available in the near extreme regime.

\begin{center}
\begin{figure}[tp]
\includegraphics[clip,width=\columnwidth]{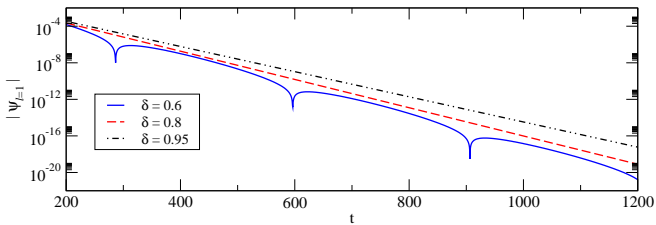} 
\caption{Scalar field evolution with $l=1$ and $r_{\star}=0$, for several
values of $\delta$. For the wormhole geometries considered, the parameters $\Lambda_{4D}=0.01$, $M=0.1$ and $Q=0.05$ ($r_{c}=17.22$) were used. }
\label{scalar_field_evolution_l1} 
\end{figure}
\end{center}

For the scalar perturbation with $l=1$, the main qualitative characteristics
of its perturbative dynamics are described as follows. If $\delta$
is close to $1$ ($C/C_{ext}$ small), the late-time decay is
(nonoscillatory) exponential,
\begin{equation}
\psi_{l=1} \sim e^{k \, t}\,\,,
\end{equation}
with $k<0$. This result is consistent with the scalar dynamics around
other asymptotically de Sitter geometries \cite{Brady-1997,Molina-2004}.
We illustrate this result in Fig.\ref{scalar_field_evolution_l1}.
For smaller values of $\delta$ (larger $C/C_{ext}$), the decay is
oscillatory, with an exponential envelope,
\begin{equation}
\psi_{l=1} \sim e^{\textrm{Im} \left(\omega_{0}^{sc}\right) \, t} \, e^{-i \,\textrm{Re} \left(\omega_{0}^{sc}\right) \, t} \,\, ,
\end{equation}
where $\omega_{0}^{sc}$ is the fundamental (lowest absolute value of the imaginary
part) quasinormal frequency associated with the $l=1$ scalar mode and $\textrm{Im} \left(\omega_{0}^{sc}\right) < 0$.
We illustrate this result in Fig. \ref{scalar_field_evolution_l1}. 

\begin{center}
\begin{figure}[tp]
\includegraphics[clip,width=\columnwidth]{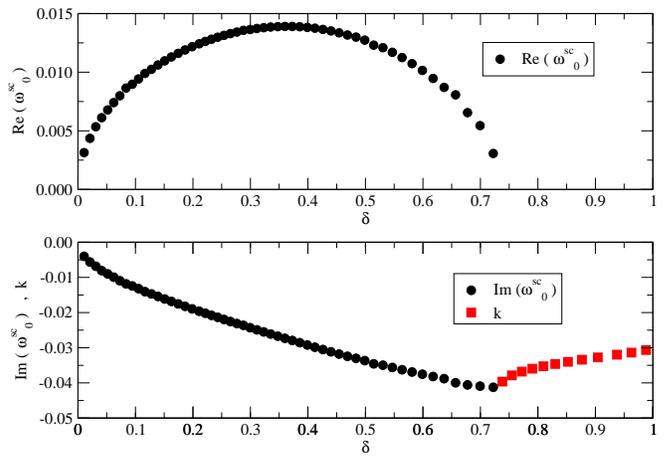} 
\caption{Dependence of the exponential coefficient $k$ and the scalar fundamental
quasinormal frequency $\omega^{sc}_{0}$ on the parameter $\delta$, for the $l=1$ mode.
For the wormhole geometries considered, the parameters $\Lambda_{4D}=0.01$, $M=0.1$ and $Q=0.05$ ($r_{c}=17.22$) were used.}
\label{k_w0_l1} 
\end{figure}
\end{center}

Typical profiles for the dependence of the parameters $k$, $\textrm{Im}\left(\omega_{0}^{sc}\right)$ and $\textrm{Re}\left(\omega_{0}^{sc}\right)$ on $\delta$ are shown
in Fig.\ref{k_w0_l1}. From these results, we see that the shift
of oscillatory and nonoscillatory modes at $t\rightarrow\infty$
is determined by the relative magnitude of $k$ and $\textrm{Im}\left(\omega_{0}^{sc}\right)$.
If $\left|k\right|>\left|\textrm{Im}\left(\omega_{0}^{sc}\right)\right|$
(small $\delta$), the nonoscillatory mode is suppressed for large
$t$, and the oscillatory phase dominates. If $\left|\textrm{Im}\left(\omega_{0}^{sc}\right)\right|>\left|k\right|$
(large enough $\delta$), the oscillatory mode is suppressed, and a late-time nonoscillatory decay is observed. 

\begin{center}
\begin{figure}[tp]
\includegraphics[clip,width=\columnwidth]{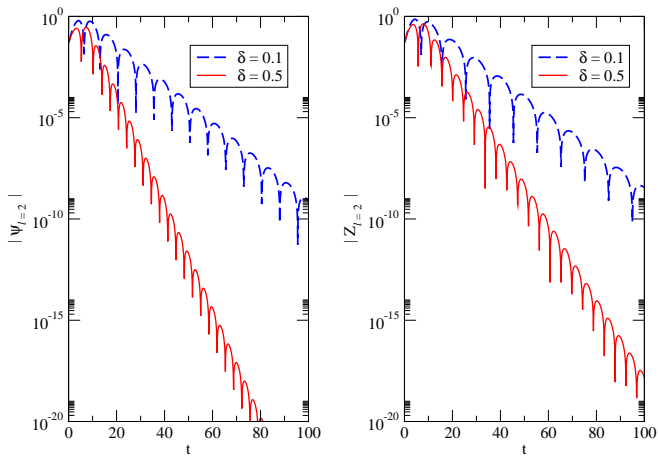} 
\caption{Scalar and gravitational perturbations with $l=2$ and $r_{\star}=0$,
for several values of $\delta$. For the wormhole geometries considered,
the parameters $\Lambda_{4D}=3$ and $M=Q=0$ ($r_{c}=1$) were used. }
\label{scalar_grav_perturbations} 
\end{figure}
\end{center}

For scalar or gravitational perturbations with $l>1$ the intermediate
and late-time dynamics is dominated by an oscillatory exponential
decay. The scalar and gravitational perturbations can be well characterized
by their fundamental quasinormal frequencies
($\omega_{0}^{sc}$ and $\omega_{0}^{grav}$ ):
\begin{equation}
\psi_{l} \sim e^{-i\,\omega_{0}^{sc}\, t} \,\, ,
\end{equation}
\begin{equation}
Z_{l} \sim e^{-i\,\omega_{0}^{grav}\, t} \,\, .
\end{equation}
These results are illustrated in Fig.\ref{scalar_grav_perturbations}.
We have not observed nonoscillatory exponential decays for the scalar
or gravitational perturbations with $l>1$, considering values of
$C/C_{ext}$ as low as $10^{-4}$ ($\delta \approx 0.999$).

\begin{table*}
\caption{Fundamental quasinormal frequencies for the scalar perturbation for
several values of $\delta$ and $l$. For the wormhole geometries
considered, the parameters $\Lambda_{4D}=3$ and $M=Q=0$ ($r_{c}=1$)
were used. Relative errors $\left(\Delta\%\right)$ for the near extreme
results are indicated.}
\label{qnm_sc}
\begin{tabular*}{\textwidth}{*{6}{c@{\extracolsep{\fill}}}}
\hline 
 &  & \multicolumn{2}{c}{Direct integration} & \multicolumn{2}{c}{Near extreme results}\\
$l$ & $\delta$ & $\textrm{Re}\left(\omega_{0}^{sc}\right)$ & $\textrm{Im}\left(\omega_{0}^{sc}\right)$ & $\textrm{Re}\left(\omega_{0}^{sc}\right)$ $\left(\Delta\%\right)$ & $\textrm{Im}\left(\omega_{0}^{sc}\right)$$\left(\Delta\%\right)$\\
\hline 
1 & 0.001 & 0.01660 & -0.02143 & 0.01659 (0.06\%) & -0.02142 (0.05\%)  \\
1 & 0.01  & 0.05224 & -0.06806 & 0.05246 (0.42\%) & -0.06773 (0.49\%)  \\
1 & 0.1   & 0.1580  & -0.2175  & 0.1659 (5.00\%)  & -0.2142 (1.52\%)   \\
1 & 0.3   & 0.2318  & -0.4130  & 0.2874 (24.0\%)  & -0.3710 (10.2\%)   \\
1 & 0.5   & 0.2084  & -0.5747  & 0.3710 (78.02\%) & -0.4789 (16.7\%)   \\
  &       &         &          &                  &                    \\
2 & 0.001 & 0.04177 & -0.02143 & 0.04175 (0.05\%) & -0.02142 (0.05\%)  \\
2 & 0.01  & 0.1321  & -0.06806 & 0.1320 (0.08\%)  & -0.06773 (0.48\%)  \\
2 & 0.1   & 0.4192  & -0.2248  & 0.4177 (0.36\%)  & -0.2143 (4.67\%)   \\
2 & 0.3   & 0.7256  & -0.4307  & 0.7232 (0.33\%)  & -0.3710 (13.8\%)   \\
2 & 0.5   & 0.9157  & -0.6025  & 0.9336 (1.95\%)  & -0.4789 (20.5\%)   \\
2 & 0.7   & 1.017   & -0.7086  & 1.105 (8.65\%)   & -0.5667 (20.0\%)   \\
2 & 0.9   & 1.012   & -0.9078  & 1.253 (23.8\%)   & -0.6426 (29.2\%)   \\
\hline
\end{tabular*}
\end{table*}

\begin{table*}
\caption{Fundamental quasinormal frequencies for the gravitational perturbation
for several values of $\delta$ and $l$. For the wormhole geometries
considered, the parameters $\Lambda_{4D}=3$ and $M=Q=0$ ($r_{c}=1$)
were used. Relative errors $\left(\Delta\%\right)$ for the near extreme
results are indicated.}

\label{qnm_grav}
\begin{tabular*}{\textwidth}{*{6}{c@{\extracolsep{\fill}}}}
\hline 
 &  & \multicolumn{2}{c}{Direct integration} & \multicolumn{2}{c}{Near extreme results}\\
$l$ & $\delta$ & $\textrm{Re}\left(\omega_{0}^{grav}\right)$ & $\textrm{Im}\left(\omega_{0}^{grav}\right)$ & $\textrm{Re}\left(\omega_{0}^{grav}\right)$ $\left(\Delta\%\right)$ & $\textrm{Im}\left(\omega_{0}^{grav}\right)$ $\left(\Delta\%\right)$\\
\hline 
2 & 0.001 & 0.03178 & -0.02141 & 0.03177 (0.03\%) & -0.02142 (0.05\%)  \\
2 & 0.01  & 0.1004  & -0.06758 & 0.1005 (0.09\%)  & -0.06773 (0.22\%)  \\
2 & 0.1   & 0.3167  & -0.2095  & 0.3177 (0.32\%)  & -0.2142 (2.24\%)   \\
2 & 0.3   & 0.5457  & -0.3354  & 0.5503 (0.84\%)  & -0.3710 (10.6\%)   \\
2 & 0.5   & 0.6962  & -0.4383  & 0.7104 (2.04\%)  & -0.4789 (9.26\%)   \\
2 & 0.7   & 0.8054  & -0.5074  & 0.8405 (4.36\%)  & -0.5667 (11.7\%)   \\
2 & 0.9   & 0.8647  & -0.5622  & 0.9531 (10.2\%)  & -0.6426 (14.7\%)   \\
  &       &         &          &                  &                   \\
3 & 0.001 & 0.05669 & -0.02142 & 0.05668 (0.02\%) & -0.02142 (0.00\%)  \\
3 & 0.01  & 0.1792  & -0.06778 & 0.1792 (0.0\%)   & -0.06773 (0.07\%)  \\
3 & 0.1   & 0.5664  & -0.2158  & 0.5667 (0.05\%)  & -0.2142 (0.74\%)   \\
3 & 0.3   & 0.9785  & -0.3803  & 0.9815 (0.31\%)  & -0.3710 (2.24\%)   \\
3 & 0.5   & 1.254   & -0.5012  & 1.267 (1.04\%)   & -0.4789 (4.45\%)   \\
3 & 0.7   & 1.458   & -0.6074  & 1.499 (2.81\%)   & -0.5667 (6.70\%)   \\
3 & 0.9   & 1.577   & -0.7013  & 1.700 (7.80\%)   & -0.6426 (8.37\%)   \\
\hline
\end{tabular*}
\end{table*}

In the near extreme regime ($0<\delta\ll1$ or $C/C_{ext}\lesssim 1$),
considered in Sec. IV, the scalar and gravitational quasinormal mode
spectra can analytically determined. Explicit analytic expressions
for the functions $V_{sc}(r(r_{\star}))$ and $V_{sc}(r(r_{\star}))$
are usually not available, except in particular limits. One of these
limits is the near extreme regime. Following an approach similar to the
one used in \cite{Cardoso-2003,Molina-2003}, the result (\ref{metric_near_extreme}) allows both effective potentials to be written as
\begin{equation}
V\left(r(r_{\star})\right) = \frac{V_{max}}{\cosh^{2} \left( \kappa_{c} \, r_{\star}\right)} \,\, ,
\label{Poschl-Teller_potential}
\end{equation}
with the surface gravity $\kappa_{c}$ presented in Eq.(\ref{kappac_near_extreme}). The constants $V_{max}$, for the scalar and gravitational perturbations
($V_{max}^{sc}$ and $V_{max}^{grav}$, respectively) are
\begin{gather}
V_{max}^{sc} = \nonumber \\
\delta \, \Lambda_{4D} \, l(l+1) \, \frac{\left(r_{c} - r_{0}\right) \left(r_{c} - r_{+}\right) \left(r_{c} - r_{-}\right) \left(r_{c} - r_{n}\right)}{3r_{c}^{4}} , \nonumber \\
\textrm{with} \,\, l>0 \,\, ,
\label{Vmax_sc}
\end{gather}
\begin{gather}
V_{max}^{grav} = \nonumber \\
\delta \, \Lambda_{4D} \, (l+2)(l-1) \, \frac{\left(r_{c} - r_{0}\right) \left(r_{c} - r_{+}\right) \left(r_{c} - r_{-}\right) \left(r_{c} - r_{n}\right)}{3r_{c}^{4}} , \nonumber \\
\textrm{with} \,\, l>1 \,\, .
\label{Vmax_grav}
\end{gather}

The potential in (\ref{Poschl-Teller_potential}) is the so-called
P\"{o}schl-Teller potential \cite{Poschl-Teller}. It has been extensively
studied, and, in particular, the quasinormal modes associated with it
have been calculated \cite{Ferrari-1984,Beyer-1999}. Using the results in \cite{Ferrari-1984,Beyer-1999}, we have for
the scalar and gravitational quasinormal mode spectra, in the near
extreme regime
\begin{equation}
\omega_{n}^{sc} = \kappa_{c} \left[ \sqrt{\frac{V_{max}^{sc}}{\kappa_{c}^{2}} - \frac{1}{4}} - \left(n + \frac{1}{2}\right) \, i \right] \,\, ,
\label{qnm_PT_sc}
\end{equation}
\begin{equation}
\omega_{n}^{grav} = \kappa_{c} \left[\sqrt{\frac{V_{max}^{grav}}{\kappa_{c}^{2}} - \frac{1}{4}} - \left(n+\frac{1}{2}\right) \, i\right] \,\, .
\label{qnm_PT_grav}
\end{equation}
with $\kappa_{c}$, $V_{max}^{sc}$ and $V_{max}^{grav}$ given by
expressions (\ref{kappac_near_extreme}), (\ref{Vmax_sc}) and (\ref{Vmax_grav})
respectively.

The fundamental ($n=0$) modes dominate the late-time decay. We stress
the excellent concordance of the analytical expressions (\ref{qnm_PT_sc})-(\ref{qnm_PT_grav}) with the numerical results in the near extreme regime. Moving away from the near extreme limit, we consider the quasinormal
spectra for higher values of $\delta$. A direct integration approach
has been used. Quasinormal frequencies for the scalar and gravitational
perturbations are presented in Tables \ref{qnm_sc} and \ref{qnm_grav}.

In all numerical calculations performed, the concordance between the
numerical and near extreme approximation improves as $\delta$ is
made smaller. This is a consistency check for the numerical results
and an indication that the near extreme results are indeed adequate
when the appropriate limit is taken. We illustrate this point in Tables
\ref{qnm_sc} and \ref{qnm_grav}. Moreover, the analytical expression in Eq.(\ref{qnm_PT_grav}) for the gravitational sector
appears to work well even when the condition $\delta \ll 1$
is not strictly satisfied, as suggested by the data presented, for
example, in Table \ref{qnm_grav}.

\section{Final remarks}

We have obtained a family of exact solutions of the effective Einstein equations in a asymptotically de Sitter Randall-Sundrum brane. This family includes naked singularities, but also solutions which describe wormholes. Maximal extensions of the solutions were studied. We have shown that the extensions describe Lorentzian, traversable, wormhole space-times which connect regions bounded by cosmological horizons. It should be noted that, although the existence of a local, asymptotically de Sitter, solution for a metric in a Randall-Sumdrum scenario might be expected, it is not obvious that there would exist solutions regular everywhere. The explicit solutions constructed here have this characteristics.

One basic requirement, if the geometries obtained are to be considered as physically relevant, is the stability of the derived geometries under first-order perturbations. We have treated this question here considering scalar and axial gravitational perturbations. An important result in the perturbative analysis performed in this work is that \emph{no unstable modes were found}. 

Moreover, the detailed numerical and analytical treatment presented sketches a picture of the perturbative dynamics. Scalar spherically symmetric modes typically decay to a nonzero constant asymptotically. This is reminiscent of a feature already observed in considerations involving de Sitter black holes \cite{Brady-1997,Molina-2004}. Although oscillatory and nonoscillatory decays bounded by exponential envelopes were observed, no power-law tails appeared, which also resembles the dynamics around asymptotically de Sitter black holes \cite{Brady-1997,Molina-2004}.

One interesting limit of the geometries derived in this work is their near extreme regime. This limit is interesting because the geometry becomes very simple, while still preserving the causal structure of the nonextreme case. In fact, in the near extreme regime the quasinormal spectra of the perturbations considered can be analytically determined, which is something not common in the literature. Moreover, the comparison between the full numerical results and the near extreme approximation shows good agreement for the fundamental overtone. We consider this result a strong argument for the validation of both approaches. Besides, the near extreme analytical results appear to describe reasonably well the gravitational quasinormal spectra even outside this limit.

\begin{acknowledgments}
This work was partially supported by Conselho Nacional de Desenvolvimento Cient\'{\i}fico e Tecnol\'ogico (CNPq) and Coordena\c{c}\~ao de Aperfei\c{c}oamento de Pessoal de N\'{\i}vel Superior (CAPES), Brazil.
\end{acknowledgments}

\end{document}